# Single layer of $MX_3$ (M=Ti, Zr; X=S, Se, Te): a new platform for nano-electronics and optics

Yingdi Jin,[a] Xingxing Li[a] and Jinlong Yang[*a,b]

A serial of two dimensional titanium and zirconium trichalcogenides nanosheets $MX_3$ (M=Ti, Zr; X=S, Se, Te) are investigated based on first-principles calculations. The evaluated low cleavage energy indicates that stable two dimensional monolayers can be exfoliated from their bulk crystals in experiment. Electronic studies reveal very rich electronic properties in these monolayers, including metallic $TiTe_3$ and $ZrTe_3$, direct band gap semiconductor $TiS_3$ and indirect band gap semiconductors $TiSe_3$, $ZrS_3$ and $ZrSe_3$. The band gaps of all the semiconductors are between 0.57~1.90 eV, which implies their potential applications in nano-electronics. And the calculated effective masses demonstrate highly anisotropic conduction properties for all the semiconductors. Optically, $TiS_3$ and $TiSe_3$ monolayers exhibit good light absorption in the visible and near-infrared region respectively, indicating their potential applications in optical devices. In particular, the highly anisotropic optical absorption of $TiS_3$ monolayer suggests it could be used in designing nano optical waveguide polarizers.

## Introduction

Since the successful isolation of graphene,[1] two dimensional (2D) materials have attracted tremendous attentions with a wide range of physicochemical properties and potential applications. Pristine graphene lacks a finite band gap, which is essential for controllable and reliable transistor operation. Thus, for nano-electronics and optics, it is necessary to explore other two dimensional materials[2-12] with suitable band gaps, for example transition metal dichalcogenides[5-7] and the recently rediscovered phosphorene.[8-12] Moreover, benefited from the state-of-art liquid exfoliation method,[13-15] it is now possible to extract single layered materials from any van der Waals (vdW) stacked layered crystals. Transition metal trichalcogenides $MX_3$, where M is transition metal Ti or Zr and X is S, Se or Te, are typical vdW stacked layered materials.[16] Therefore, it is expected that 2D $MX_3$ nanosheets can be obtained by exfoliating their bulk counterparts.

Experimentally, the electrical and optical properties of bulk $TiS_3$, $ZrSe_3$, $HfSe_3$, $ZrS_3$ and $ZrTe_3$ have been studied.[17-21] Few-layer $TiS_3$ nanoribbons have been successfully isolated and the macroscopic films of $TiS_3$ ribbons show a direct band gap of ~1.1 eV and ultrahigh photoresponse.[22-24] However, in theoretical respect, previous studies of these materials were limited to the band structure calculations of bulk $ZrSe_3$ and $ZrTe_3$.[25-28] Until very recently, the electronic structure of $TiS_3$ monolayer was computed, and an indirect to direct band gap transition from bulk to monolayer was found.[29] Before further applying these materials in nano-electronic and optical devices, a more comprehensive study from theoretical aspect would be needed.

In this paper, we systematically study the structural, electronic and optical properties of monolayer $MX_3$ sheets. The cleavage energies of bulk $MX_3$ are evaluated to be close to that of graphite, directly demonstrating the feasibility of obtaining 2D $MX_3$ crystals by exfoliation. Electronic structure calculations show that the trisulfide and triselenide monolayers are semiconductors with band gaps in the range of 0.57-1.90 eV, while the two tritelluride monolayers are metallic. Contrary to $TiS_3$, the $TiSe_3$, $ZrS_3$ and $ZrSe_3$ monolayers are all indirect gap semiconductors as their bulks and the indirect to direct band gap transition no longer appears. We further find that due to structural anisotropy, all the semiconductors possess highly anisotropic effective masses and conductive properties. Compared to transition metal dichalcogenides, the extra X-X bonds in these trichalcogenides monolayers introduce states far below the Fermi level and tend not to affect the electronic properties of $MX_3$ significantly. Finally, the good visible/near-infrared light absorption of $TiS_3$/$TiSe_3$ monolayer implies their potential applications in nano-optical devices. In addition, the high anisotropy in light absorption of $TiS_3$ monolayer offers a possibility to fabricate optical waveguide polarizers.

[a] *Hefei National Laboratory of Physical Science at the Microscale, University of Science and Technology of China, Hefei, Anhui 230026, China, E-mail: jlyang@ustc.edu.cn; Fax:+86 551 63603748; Tel:+86 551 63606408*
[b] *Synergetic Innovation Center of Quantum Information & Quantum Physics, University of Science and Technology of China, Hefei, Anhui 230026, China*

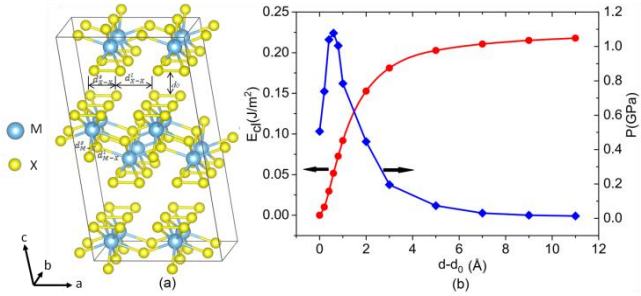

**Figure 1.** (a) Crystal structure of MX$_3$ in the ZrSe$_3$-type structure. Yellow and light blue spheres refer to X (X=S, Se and Te) and M (M=Ti, Zr) atoms. $d^s_{M-X}$, $d^l_{M-X}$ denote the short and long M-X bonds, $d^s_{X-X}$, $d^l_{X-X}$ denote the short and long X-X bonds, respectively. (b) Cleavage energy E$_{cl}$ (red) and cleavage strength P (blue) as a function of the separation d between two fractured monolayers.

**Table 1.** Summary of structural parameters and ideal cleavage cohesion energy (E$_{cc}$) for monolayer MX$_3$. ($a$, $b$, $\beta$), $d^s_{M-X}$, $d^l_{M-X}$, $d^s_{X-X}$, $d^l_{X-X}$ are the lattice constants, short and long M-X bonds, short and long X-X bonds, respectively.

|  | TiS$_3$ | TiSe$_3$ | TiTe$_3$ | ZrS$_3$ | ZrSe$_3$ | ZrTe$_3$ |
|---|---|---|---|---|---|---|
| $a$ (Å) | 4.993 | 5.328 | 5.938 | 5.173 | 5.450 | 5.968 |
| $b$ (Å) | 3.393 | 3.538 | 3.729 | 3.612 | 3.740 | 3.947 |
| $\beta$ (°) | 97.29 | 97.56 | 97.55 | 97.35 | 97.54 | 97.77 |
| $d^s_{M-X}$ (Å) | 2.491 | 2.602 | 2.826 | 2.622 | 2.761 | 2.952 |
| $d^l_{M-X}$ (Å) | 2.653 | 2.843 | 3.324 | 2.732 | 2.892 | 3.200 |
| $d^s_{X-X}$ (Å) | 2.038 | 2.344 | 2.821 | 2.073 | 2.377 | 2.855 |
| $d^l_{X-X}$ (Å) | 2.956 | 2.984 | 3.120 | 3.101 | 3.073 | 3.113 |
| E$_{cc}$ (J/m$^2$) | 0.226 | 0.376 | 0.704 | 0.240 | 0.373 | 0.677 |

## Computational Methods

Geometrical optimizations and electronic structure calculations are performed by using the density functional method (DFT) implemented in the Vienna ab initio Simulation Package (VASP).[30] The exchange-correlation energy is treated by using the Perdew-Burke-Ernzerhof (PBE) functional, and the Grimme's DFT-D2 dispersion correction[31] is applied to account for the long-range van der Waals interactions. Since the PBE functional tends to underestimate the band gap of semiconductors, the hybrid HSE06 functional[32] is then adopted to get accurate band gaps of single layer MX$_3$. A vacuum space of ~20 Å along the direction normal to the monolayer plane is used so that the interlayer interaction generated by the periodic boundary condition can be neglected. The ion-electron interaction is treated by using the projector-augment-wave (PAW) technique.[33,34] For geometrical optimization, both lattice constants and atomic positions are relaxed until the forces on the atoms are less than 0.02 eV/Å and the total energy change is less than $1.0 \times 10^{-5}$ eV. A 7×10×1 Monkhorst-Pack[35] grid and a kinetic energy cutoff of 500 eV are selected. For static calculations, a finer 14×20×1 grid is chosen.

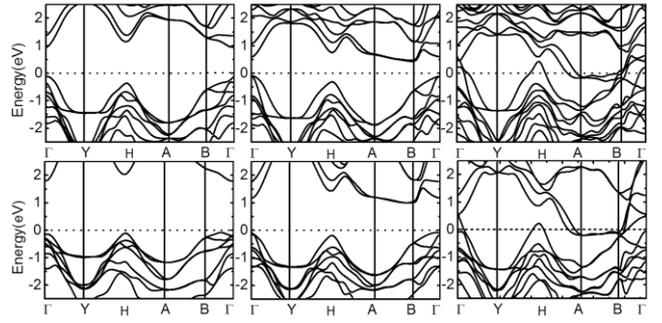

**Figure 2.** Electronic band structure calculated with HSE06 functional for (a) TiS$_3$, (b) TiSe$_3$, (c) TiTe$_3$, (d) ZrS$_3$, (e) ZrSe$_3$ and (f) ZrTe$_3$ monolayers. The Fermi level is set to zero. Γ (0.0, 0.0, 0.0), Y (0.0, 0.5, 0.0), H (0.25, 0.0, 0.0), A (0.5, -0.5, 0.0), B (0.5, 0.0, 0.0) refer to the special points in the first Brillouin zone.

## Results and Discussion

We start our calculations from the benchmarks of the geometrical and electronic properties of bulk TiS$_3$ to evaluate the accuracy of the methods adopted in this work. The optimized lattice constants of bulk TiS$_3$ by PBE with Grimme's D2 correction is $a$= 4.982 Å, $b$= 3.392 Å, $c$= 8.887 Å and $\beta$= 97.24°, matching well with the experimental values ($a$= 4.958 Å, $b$= 3.401 Å, $c$= 8.778 Å and $\beta$= 97.32°).[36] The indirect and direct band gaps of bulk TiS$_3$ calculated with HSE06 functional are 1.02, 1.13 eV respectively. The latter is consistent with the experimentally observed optical band gap of bulk TiS$_3$ (~ 1 eV[17, 37]), confirming the validity of our methods.

We then turn to the structural properties of MX$_3$ monolayers. Bulk MX$_3$ usually crystallize in the monoclinic ZrSe$_3$-type structure composed of vdW stacked parallel sheets, where one-dimensional chains of triangular MX$_3$ prisms along the **b** lattice direction are linked together in the **a** direction [Figure 1(a)]. Thus, monolayer MX$_3$ can be viewed as inter-connected one-dimensional chains of triangular MX$_3$ prisms, with the M-X bonds in the chains being significantly shorter than those between the chains (Table 1). Moreover, different from transition metal dichalcogenides, the structure of MX$_3$ possesses two types of chalcogen atoms, i.e. outermost X atoms and inner X atoms. The outermost X atoms form X-X chains with alternating bond length $d^l_{X-X}$ and $d^s_{X-X}$, displaying in Figure 1(a). The structural parameters of the optimized MX$_3$ monolayers are summarized in Table 1. Clearly we can see the lattice constants expand with the increase of atomic radius, for example, since the atomic radius of Te (Zr) is bigger than that of S (Ti), the lattice constants $a$ and $b$ of TiTe$_3$ (ZrS$_3$) are about 18.9% (3.6%) and 9.9% (6.5%) larger than those of TiS$_3$.

In order to investigate the possibility of obtaining 2D MX$_3$ monolayers from their bulk crystals, we calculate the ideal cleavage cohesion energy of each material by introducing a fracture in the bulk.[38] The variations of total energy according to the separation $d$ between the fractured parts are calculated to simulate the exfoliation procedure. As shown in Figure 1(b), the total energy of TiS$_3$ increases with $d$ and converges to its ideal cleavage cohesion energy of 0.23 J/m$^2$ with a cleavage strength of 1.1 GPa. In the same way, the ideal cleavage cohesion energies of other materials are also evaluated and listed in the last line of Table 1. Here, both the ideal cleavage cohesion energies of $TiS_3$ and $ZrS_3$ are smaller than the experimentally estimated cleavage energy in graphite (~0.36 J/m$^2$), and the values of $TiSe_3$ and $ZrSe_3$ are very close to that of graphite, which directly demonstrate the feasibility of exfoliation. As with $TiTe_3$ and $ZrTe_3$, which possess about twice the cleavage energy of graphite, the exfoliation process would be harder compared to the trisulfides and triselenides.

**Table 2.** Summary of electronic structure of bulk and monolayer $MX_3$. The capitals I, D, M indicate the material is an indirect or direct band gap semiconductor, or metal, respectively. Both the values of indirect and direct band gaps $E_g^I$ and $E_g^d$ are calculated. $M_e^*$ and $M_h^*$ are the effective masses of electron and hole along **a** and **b** directions at conduction band minimum and valence band maximum, respectively. The unit is of the electron rest mass ($m_{e0}$). The density of states (DOS) per chemical formula at the Fermi level for metallic $TiTe_3$ and $ZrTe_3$ are also given.

|  | $TiS_3$ | $TiSe_3$ | $TiTe_3$ | $ZrS_3$ | $ZrSe_3$ | $ZrTe_3$ |
|---|---|---|---|---|---|---|
| Bulk | I | I | M | I | I | M |
| $E_g^I$(eV) | 1.02/ΓZ | 0.21/ΓB | - | 1.83/HZ | 0.75/ΓB | - |
| $E_g^d$(eV) | 1.13/ΓΓ | 0.73/ΓΓ | - | 2.13/HH | 1.29/ΓΓ | - |
| Monolayer | D | I | M | I | I | M |
| $E_g^I$(eV) | - | 0.57/ΓB | - | 1.90/HΓ | 1.17/ΓB | - |
| $E_g^d$(eV) | 1.06/ΓΓ | 0.70/ΓΓ | - | 1.96/ΓΓ | 1.28/ΓΓ | - |
| $M_e^*(m_{e0})$ |  |  |  |  |  |  |
| **a** | 1.47 | 0.19 | - | 1.30 | 0.16 | - |
| **b** | 0.41 | 4.29 | - | 0.40 | 6.72 | - |
| $M_h^*(m_{e0})$ |  |  |  |  |  |  |
| **a** | 0.32 | 3.57 | - | 1.28 | 2.36 | - |
| **b** | 0.98 | 0.85 | - | 0.42 | 0.89 | - |
| DOS($E_f$) | - | - | 0.91 | - | - | 1.01 |

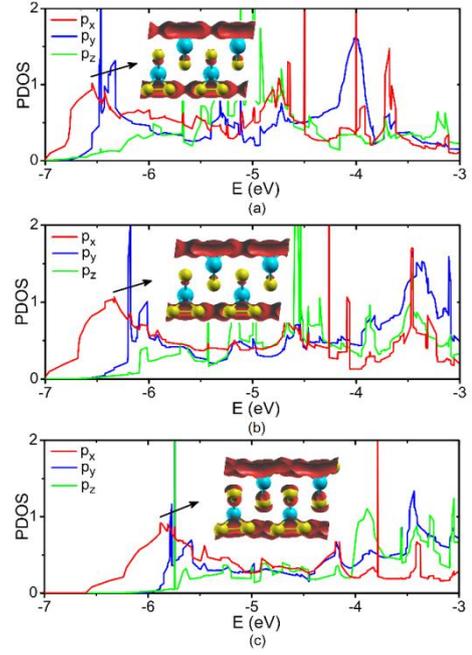

**Figure 3.** Orbital projected density of states (PDOS) of outermost X atoms in (a) $TiS_3$, (b) $TiSe_3$ and (c) $TiTe_3$ monolayers. The insets are spatial charge plots of the states (viewed along **b** axis) in energy ranges of -7.0 ~ -6.0 eV, -6.8 ~ -5.8 eV, -6.3 ~ -5.3 eV for $TiS_3$, $TiSe_3$ and $TiTe_3$ monolayers, respectively. Isovalues are set to be 0.03 e/Å$^3$. Fermi levels are all set to zero.

To study the electronic properties of $MX_3$ monolayers, the band structures are calculated, as plotted in Figure 2. The obtained band gaps are summarized in Table 2, along with those of bulk crystals for comparison. As we can see, there are three interesting features. First, an indirect to direct band gap transition from bulk to monolayer is found for $TiS_3$, in accord with the previous study.[29] Second, the band gap decreases with the increasing size of the chalcogen atoms, for example, compared with $TiS_3$, monolayer $TiSe_3$ possesses a reduced band gap of 0.57 eV, while monolayer $TiTe_3$ becomes metallic. The same trend is also observed in zirconium trichalcogenides. Last, compared with their bulk counterparts, the indirect band gaps of all monolayers increase, while the direct band gaps decrease. Thus the differences between indirect and direct band gaps of the monolayers become much smaller than the bulk. For example, the disparity between indirect and direct band gaps is 0.3 eV for bulk $ZrS_3$, while it is reduced to 0.06 eV for monolayer $ZrS_3$. However, the indirect to direct band gap transition from bulk to monolayer no longer appears.

The effective masses are indicative of the conduction properties of semiconductors. Low effective mass corresponds to high mobility of the electrons/holes and consequently high conductivity. To investigate the conduction properties of semiconducting $MX_3$ monolayers, we compute the electron and hole effective masses at conduction band minimum (CBM) and valence band maximum (VBM) according to the following equation:

$$\frac{1}{m^*(\bar{k})} = \frac{1}{\hbar^2}\frac{\partial^2 E(\bar{k})}{\partial \bar{k}^2}$$

The results are summarized in Table 2. The obtained effective masses are comparable to that of monolayer $MoS_2$ ($M_e^*= 0.48m_{e0}$, $M_h^*= 0.64m_{e0}$).[39] Due to different structural and bonding characters in the **a** and **b** lattice directions, all the materials possess highly anisotropic effective masses for both electron and hole, and the effective masses of triselenide monolayers exhibit larger anisotropy than those of trisulfide monolayers. What's more, except for $ZrS_3$, the electron and hole effective masses show entirely different anisotropy, i.e. electrons and holes prefer to conduct in different directions. Another interesting finding is that the trisulfides $TiS_3$ and $ZrS_3$ have small electron mass in the direction along the chains of triangular $MX_3$ prisms, namely **b** direction, while for the triselenides $TiSe_3$ and $ZrSe_3$, electrons are more easily transported in the **a** direction, which is nearly perpendicular to **b**.

For metallic $MTe_3$ monolayers, the electrical conductivity is directly proportional to the density of states at Fermi level N($E_F$). The calculated N($E_F$) of $TiTe_3$ and $ZrTe_3$ per chemical formula is 0.91 and 1.01, respectively (Table 2). Comparing with 0.72 states per formula of superconducting $MgB_2$,[40] the N($E_F$) values of $TiTe_3$ and $ZrTe_3$ imply the possibility to achieve 2D superconductivity in these two materials.

Comparing with monolayer transition metal dichalcogenides, monolayer $MX_3$ trichalcogenides have an additional X-X (X= S, Se, Te) bond. In order to study the effect of this additional X-X bond on the electronic properties, we calculate the orbital-projected density of states of outermost X atoms (Figure 3). One can identify that it is the $p_x$ orbital of outermost X atom that contributes to the X-X bonding states, which are far away from the Fermi level. Thus, the additional X-X bonds tend not to affect the electronic properties of MX$_3$ significantly.

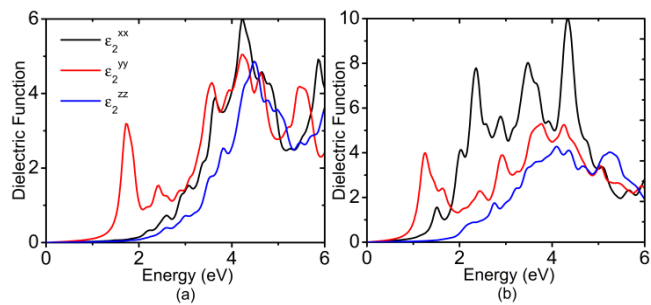

**Figure 4.** Calculated imaginary part of the frequency-dependent dielectric function $\varepsilon_2(E)$ for (a) TiS$_3$ and (b) TiSe$_3$ monolayers with HSE06 functional.

The fact that monolayer TiS$_3$ has a direct band gap of 1.06 eV, which is very close to the crystalline silicon (~1.1 eV, indirect), indicates that besides potential applications in nano-electronics, monolayer TiS$_3$ can be also very useful in optical applications, for example, as a solar absorber material. To investigate this possibility, we calculate the imaginary part of the frequency-dependent dielectric function from the summation over pairs of occupied and empty states without considering the local field effects.[41] As shown in Figure 4(a), the imaginary part of the dielectric function shows an anisotropic feature where $\varepsilon_2^{xx} \neq \varepsilon_2^{yy} \neq \varepsilon_2^{zz}$. In particular, $\varepsilon_2^{yy}$ has a very strong peak at about 1.7 eV, which is due to the interband transition between VB and CB. Because of the dispersion of VB and CB, the optical absorption lasts up to about 2.0 eV. There are also other absorption peaks due to other interband transitions, for example at around 2.4 eV, 3.0 eV and 3.5 eV. Similarly, the dielectric function of TiSe$_3$ is also computed and shows a strong absorption peak at about 1.2 eV, and other peaks at around 2.4 eV and 3.6 eV [Figure 4(b)]. These results indicate that monolayer TiS$_3$ is a promising material for absorbing visible light, while TiSe$_3$ is a good absorber for both near-infrared and visible light.

Remarkably, $\varepsilon_2^{xx}$ of TiS$_3$ monolayer is nearly zero at about 1.7 eV while the peak of $\varepsilon_2^{yy}$ is very high [Figure 4(a)]. This property implies TiS$_3$ monolayer is a candidate material for designing an optical waveguide polarizer. Polarizer is one of the most important devices in optics, which only transmits the light in one polarization direction by absorbing or reflecting the light in the other polarization direction.[42] The high anisotropy in optical absorption of TiS$_3$ monolayer indicates a high polarization sensitivity, and provides a good opportunity for designing optical waveguide polarizers.

## Conclusions

In conclusion, based on first-principles calculations, we systematically investigate the structural, electronic and optical properties of titanium and zirconium trichalcogenides monolayer sheets, which can be realized by exfoliating their bulk crystals, as suggested by our estimated low cleavage energies. Similar to MoS$_2$, TiS$_3$ undergoes an indirect to direct band gap transition from bulk to monolayer, while TiSe$_3$, ZrS$_3$ and ZrSe$_3$ monolayers have indirect band gaps in the range of 0.57 eV-1.90 eV, and the TiTe$_3$ and ZrTe$_3$ monolayers are metallic. The calculated electron and hole effective masses show that all the semiconductors have anisotropic conductive properties. Compared to transition metal dichalcogenides, the extra X-X bonds existed in these trichalcogenides monolayers do not affect the electronic properties significantly. Optical studies reveal that TiS$_3$ and TiSe$_3$ monolayers have good optical absorption in the visible and near-infrared region respectively, making them promising in fabricating nano-optical devices. Particularly, the high anisotropy in the light absorption of TiS$_3$ monolayer makes it a potential material for the design of optical waveguide polarizers.

## AUTHOR INFORMATION

**Corresponding Author**

*jlyang@ustc.edu.cn

## ACKNOWLEDGMENT

This work is partially supported by the National Key Basic Research Program (2011CB921404), by NSFC (21421063, 91021004, 21233007), by Chinese Academy of Sciences (CAS) (XDB01020300), and by USTCSCC, SCCAS, Tianjin, and Shanghai Supercomputer Centers.